\documentstyle[epsf,epsfig,wrapfig,e-e-ijmpa,twoside]{article}

\renewcommand{\thefootnote}{\fnsymbol{footnote}}	

\def\pole{{\cal P}_e}

\def\polg{{\cal P}_\gamma}
\def\rar{\rightarrow}
\def\lar{\leftarrow}

\begin{document}
\pagestyle{plain}
\thispagestyle{empty}
\normalsize\textlineskip

\begin{flushright}
{\small
SLAC--PUB--7744\\
February, 1998\\}
\end{flushright}

\vspace{.8cm}

\begin{center}
{\large\bf
COMPTON POLARIMETRY AT A 1 TEV COLLIDER\footnote{Work supported by
Department of Energy contract  DE--AC03--76SF00515.}}

\vspace{1cm}

M. WOODS\\
Stanford Linear Accelerator Center \\
Stanford University,
Stanford, CA  94309\\
\end{center}

\vfill

\begin{center}
{\large\bf
Abstract }
\end{center}

\begin{quote}
An electron beam polarization of 80$\%$ or greater will be a key feature
of a 1 TeV Linear Collider.  Accurate measurements of the beam polarization
will therefore be needed.  We discuss design considerations and 
capabilities for a Compton-scattering polarimeter located in the extraction
line from the Interaction Point.  Polarization measurements with $1\%$
accuracy taken parasitic to collision data look feasible, but  
detailed simulations are needed.  
Polarimeter design
issues are similar for both electron-positron and electron-electron collider
modes, though beam disruption creates more difficulties for the 
electron-electron mode.
\end{quote}

\vfill

\begin{center}
Presented at \\
\vspace{3mm}
{\it \boldmath$e^-e^- 97$ \\
2nd International Workshop on Electron-Electron Interactions at TEV Energies \\
September 22-24, 1997 \\
UC Santa Cruz, Santa Cruz, CA, USA}  \\
\vspace{3mm}
and \\
\vspace{3mm}
{\it {\boldmath$LC97$} \\
VII International Workshop on Linear Colliders \\
September 29 - October 3, 1997 \\
Zvenigorod, Russia} \\
\end{center}

\setcounter{footnote}{0}
\renewcommand{\thefootnote}{\alph{footnote}}

\vspace*{1pt}\textlineskip

\pagebreak	
\section{Introduction}

A Compton polarimeter analyzes either the scattered beam electrons or the
back-scattered gamma rays from the collision of a longitudinally polarized
electron beam with a circularly polarized laser beam.  The cross-section for
this process\cite{Comptxsect} is given by:
\begin{eqnarray}
\sigma_C   & = & \sigma_C^0 + \pole \polg \sigma_C^1 
\label{eq:compton}
\end{eqnarray}
where $\sigma_C^0$ is the unpolarized cross-section and $\sigma_C^1$ is the
polarized cross-section; $\pole$ is the electron linear polarization and 
$\polg$ is the laser circular polarization.
The electron beam polarization can then be deduced from measurements of the
relative Compton-scattering rates for the J=3/2 (electron and photon spins
parallel) and J=1/2 (electron and photon spins anti-parallel) initial states,
given accurate determinations of the laser polarization and the detector
analyzing power.  This is reflected in Equation~(\ref{eq:asym}):
\begin{eqnarray}
A^{meas}= {R(\rar\rar)-R(\rar\lar) \over R(\rar\rar)+R(\rar\lar)} =
         \pole\polg A_C^{det} 
\label{eq:asym}
\end{eqnarray}
where
$A_C^{det}$ is the detector analyzing power 
determined from Equation~(\ref{eq:compton}) and the detector acceptance.

$A^{meas}$ is maximal at the kinematic edge corresponding to $180^{\circ}$
backscatter in the center of mass frame,\footnote{The energy of the 
Compton-scattered electron has a minimum in the laboratory frame at the 
kinematic edge.} \hspace{1mm}
and is zero for $90^{\circ}$
scattering.  An electron polarimeter typically determines the beam
polarization from asymmetry measurements at the kinematic edge, while a gamma 
polarimeter measures the energy flow asymmetry integrated over the full 
gamma spectrum.  

\section{The Compton Polarimeter at the SLAC Linear Collider (SLC)}
\begin{wrapfigure}{r}{7cm}
\epsfig{figure=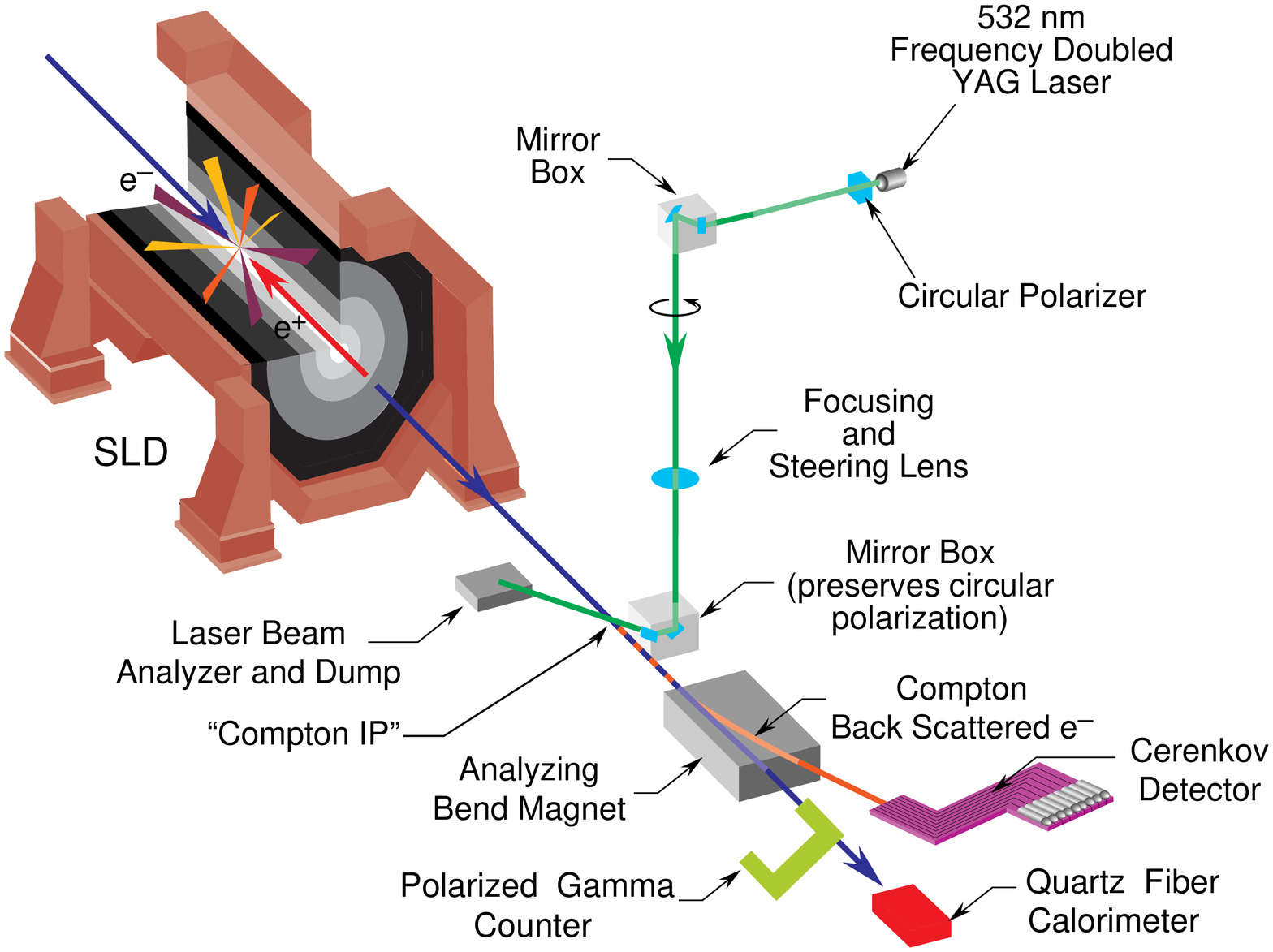,width=7cm}
\fcaption{Compton Polarimeter for the SLD Experiment}
\label{fig:sldcompton}
\end{wrapfigure}

We begin by considering the performance of the Compton polarimeter for the SLC,
and the experience from its operations during the SLD experiment.
This polarimeter,\cite{Woods1} shown in Figure~\ref{fig:sldcompton}, detects 
both Compton-scattered electrons and Compton-backscattered gammas 
from the collision of the longitudinally polarized 45.6 GeV electron
beam\cite{Woods2} with a circularly polarized photon beam. The photon beam is 
produced
from a pulsed Nd:YAG laser with a wavelength of 532 nm.  The laser is pulsed
once for every 7 electron beam pulses.\footnote{Once every 7 seconds, 
the laser is pulsed on the 6th electron pulse rather than the 7th electron 
pulse to avoid any synching of the laser pulse with instabilities in the
electron beam.  The electron beam pulse rate is 120 Hz.}  \hspace{1mm}
Laser off pulses
are used to measure backgrounds in the polarimeter detectors.
After the Compton
Interaction Point (CIP), the electrons and backscattered gammas pass through 
a dipole spectrometer.  A nine-channel threshold Cherenkov detector (CKV) 
measures electrons in the range 17 to 30 GeV.\footnote{The kinematic edge with
maximal asymmetry is at an electron energy of 17.4 GeV, and the zero-asymmetry
point is at 25.2 GeV.} \hspace{1mm}
Two detectors, a single-channel
Polarized Gamma Counter (PGC)\cite{PGC}
 and a multi-channel Quartz Fiber Calorimeter (QFC),\cite{QFC}
are located in the neutral beamline to measure the counting rates of 
the Compton-scattered gammas.  

The CKV, PGC and QFC are all 
threshold Cherenkov detectors, which are readout by photomultiplier tubes and 
charge-sensitive ADCs.  The CKV detector uses propane as the Cherenkov 
radiator, while the PGC uses ethylene and the QFC uses quartz fibers.  
Table~\ref{tab:detector}
summarizes the index of refraction and Cherenkov threshold energy for each 
of these detectors.  A high threshold energy is desirable to discriminate
against background sources of gammas from synchrotron radiation produced in
the dipole spectrometer and beamstrahlung produced in the collision process.
In fact, the PGC and QFC located in the neutral beamline only make 
polarization measurements when the beams are not in collision due to the large
beamstrahlung backgrounds.  Typical signal and background sources of gammas
are summarized in Table~\ref{tab:rates}.\footnote{These rates correspond to 
running 
conditions during the 1997 SLD run with beam intensities of $4.0 \cdot 10^{10}$
per pulse and luminosities of $1.5 \cdot 10^{30} cm^{-2}s^{-1}$.}  
\hspace{1mm} The dipole 
spectrometer is composed of a 
soft bend (1 mrad bend angle) and a hard bend (17 mrad bend angle), and the
synchrotron radiation from each is listed separately.  The soft bend magnet
does not produce a significant background source, but the hard bend magnet 
does and especially so for the QFC detector.  There is a separation at the
QFC, however, between the swath of hard bend synchrotron radiation
and the Compton gammas due to the presence of the soft bend.  The geometry
of the detector takes advantage of this, and with careful work on shielding
against scattered hard bend gammas from flanges and beampipes, acceptable
backgrounds can be achieved in the QFC.  Shielding against gammas scattered
from beampipes, flanges and apertures is also important for the PGC and 
CKV detectors.  Additionally, the CKV detector makes 
polarization measurements during beam collisions; it must be shielded against
beamstrahlung radiation and
tails of the disrupted electron beam which can hit apertures near the 
detector.  It is necessary to heavily shield these detectors and their
phototubes with lead.  
Typical Compton signal:background 
levels in the polarimeter detectors are 2:1 in the CKV detector during 
collisions, 10:1 in the PGC detector during electron-only running, and 1:1
in the QFC detector during electron-only running. 

\begin{table} [tbp]
\tcaption{Threshold Cherenkov Counters}
\vspace{2mm}
\begin{center}
\begin{tabular}{|c|clc|}
\hline
Detector	& Material	& Index of	& Threshold Energy	\\
		&		& Refraction 	&			\\
\hline
CKV		& Propane	& 1.0011	& 11 MeV		\\
PGC		& Ethylene	& 1.0007	& 14 MeV		\\
QFC		& Quartz	& 1.5		& 0.2 MeV		\\
\hline
\end{tabular}
\end{center}
\label{tab:detector}
\end{table}

\begin{table} [tbp]
\tcaption{Neutral Beam Gammas}
\vspace{2mm}
\begin{center}
\begin{tabular}{|c|crr|}
\hline
Gamma Source	& Gammas/Pulse	& $<E_{\gamma}>$	& $E_{TOTAL}$ 	\\
\hline
Compton		& 1000		& 15 GeV		& 15 TeV	\\
Soft Bend	& $4 \cdot 10^{10}$ 	& 15 keV	& 600 TeV	\\
Hard Bend	& $6.7 \cdot 10^{11}$	& 0.5 MeV	& $3.3 \cdot 
10^5$ TeV \\
Beamstrahlung	& $3.6 \cdot 10^{10}$	& 30 MeV	& $1.1 \cdot 10^6$ TeV
\\
\hline
\end{tabular}
\end{center}
\label{tab:rates}
\end{table}

Because the CKV detector is the only one which can make polarization 
measurements during beam collisions, it is the primary detector and the
most carefully analyzed.  A summary of the systematic errors associated
with its polarization measurement is given in Table~\ref{tab:systematic}.  
The error noted for the SLC IP vs Compton IP reflects that the 
luminosity-weighted polarization at the SLC IP can differ from the average
polarization measured at the Compton IP.  This 
includes the effects of depolarization from the
beam-beam interaction, chromatic effects, and steering effects.\cite{Woods2} 
Analysis of data from the PGC and QFC detectors is not yet complete, but 
preliminary results are consistent with the CKV result to within 1$\%$.
Typical beam polarizations for the SLD experiment have been in the range
$74-78\%$.

\begin{table} [tbp]
\tcaption{Systematic Errors for CKV Polarization Measurement}
\vspace{2mm}
\begin{center}
\begin{tabular}{|c|c|}
\hline
Item		&	Error	\\
\hline
Analyzing Power		&	$0.3\%$	\\
Detector Linearity	&	$0.5\%$	\\
Electronics Linearity	&	$0.2\%$	\\
Laser Polarization	&	$0.2\%$	\\
SLC IP vs Compton IP	&	$0.2\%$	\\
\hline
Total			&	$0.7\%$	\\
\hline
\end{tabular}
\end{center}
\label{tab:systematic}
\end{table}

\section{Polarimetry at a 1 TeV Linear Collider (TLC)}

The primary polarimeter at a TLC should be a Compton
polarimeter in the extraction line from the IP.  Additionally, there should
be a Mott polarimeter at the polarized electron source and a polarimeter at
or following the Damping Ring.  

\begin{figure}
\begin{center}
\epsfig{figure=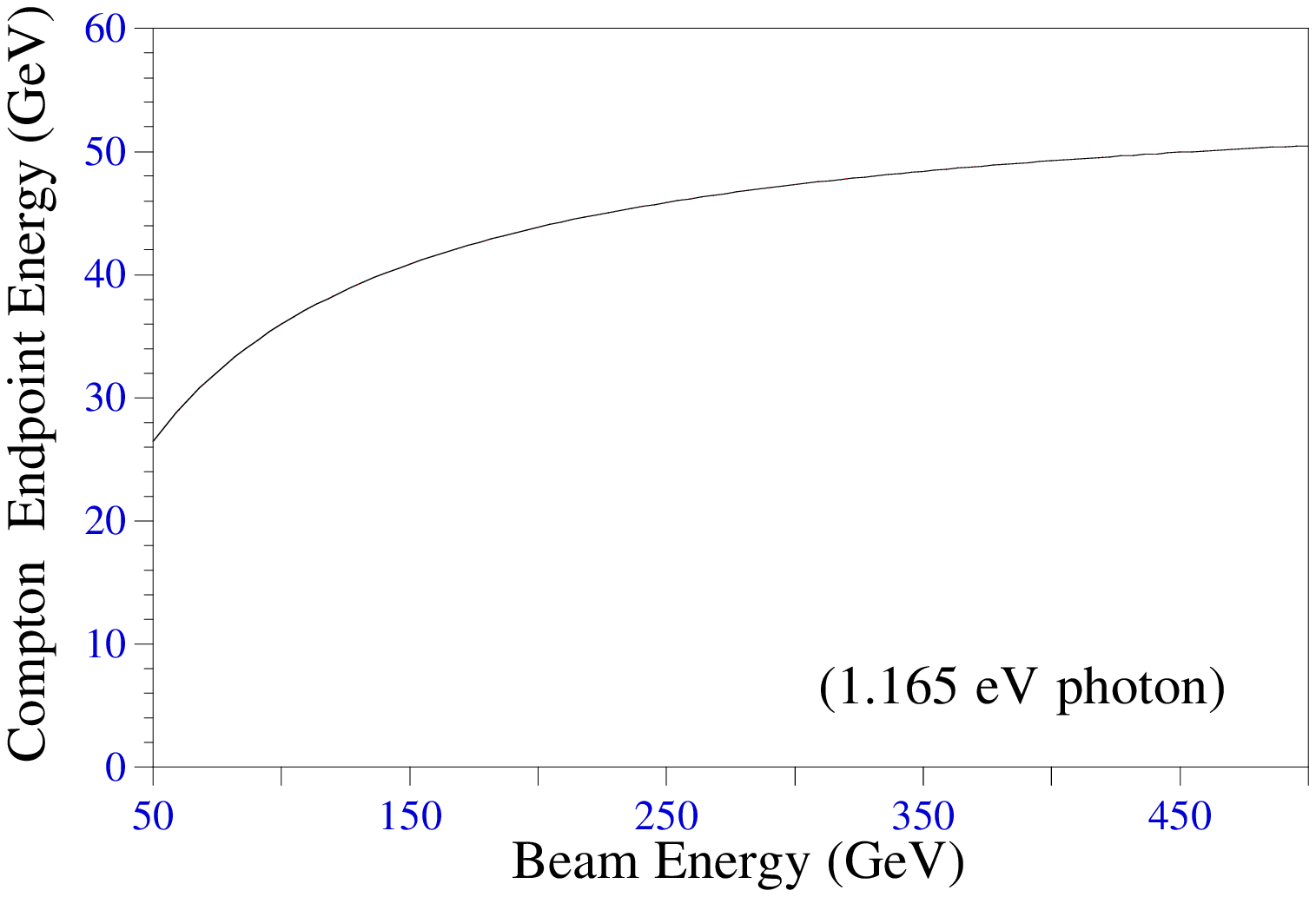,width=6.2cm}\hfill
\epsfig{figure=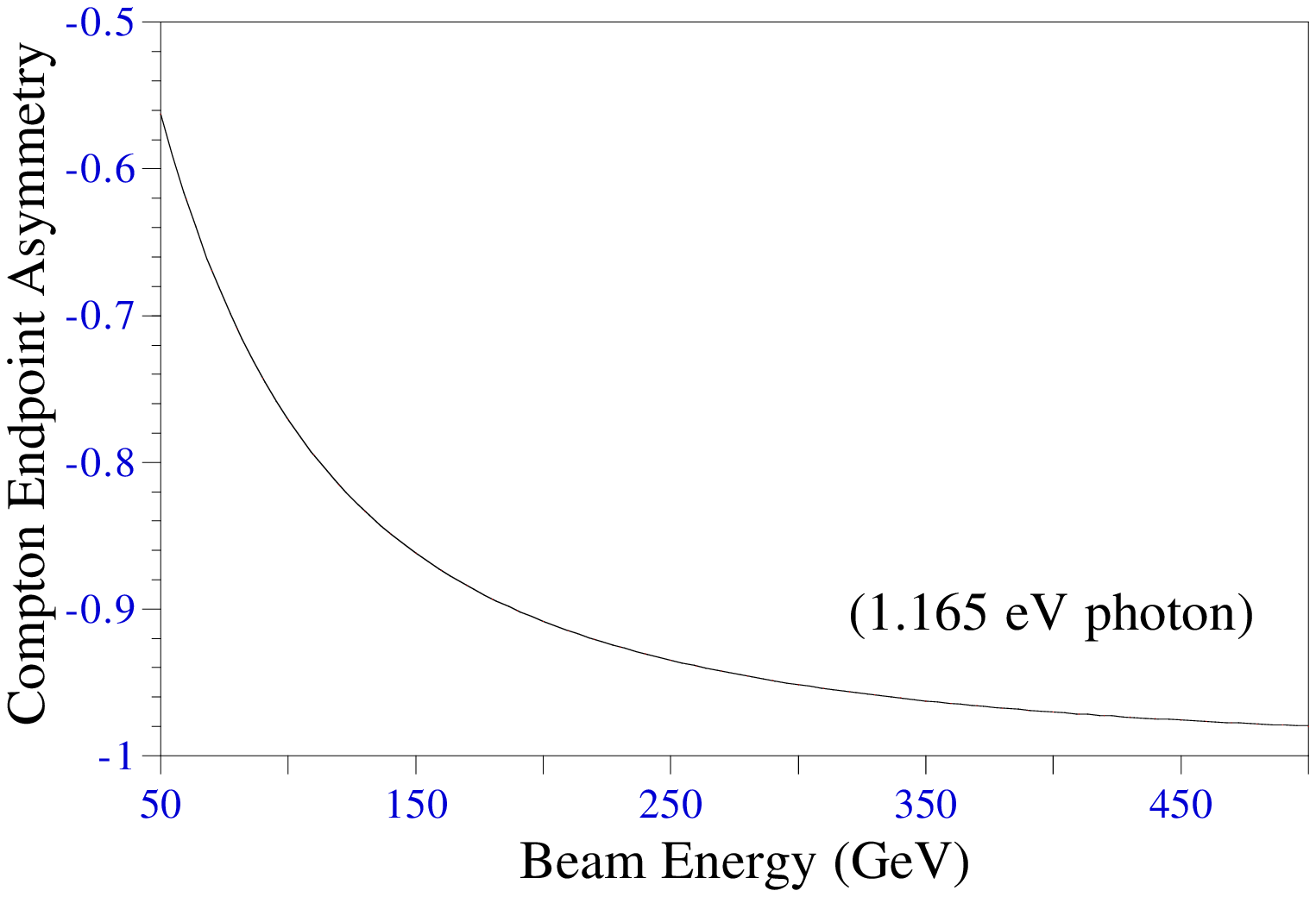,width=6.2cm}
\end{center}
\fcaption{Compton endpoint energy and endpoint asymmetry vs beam energy}
\label{fig:endpoint}
\end{figure}

\begin{wrapfigure}{r}{6.2cm}
\epsfig{figure=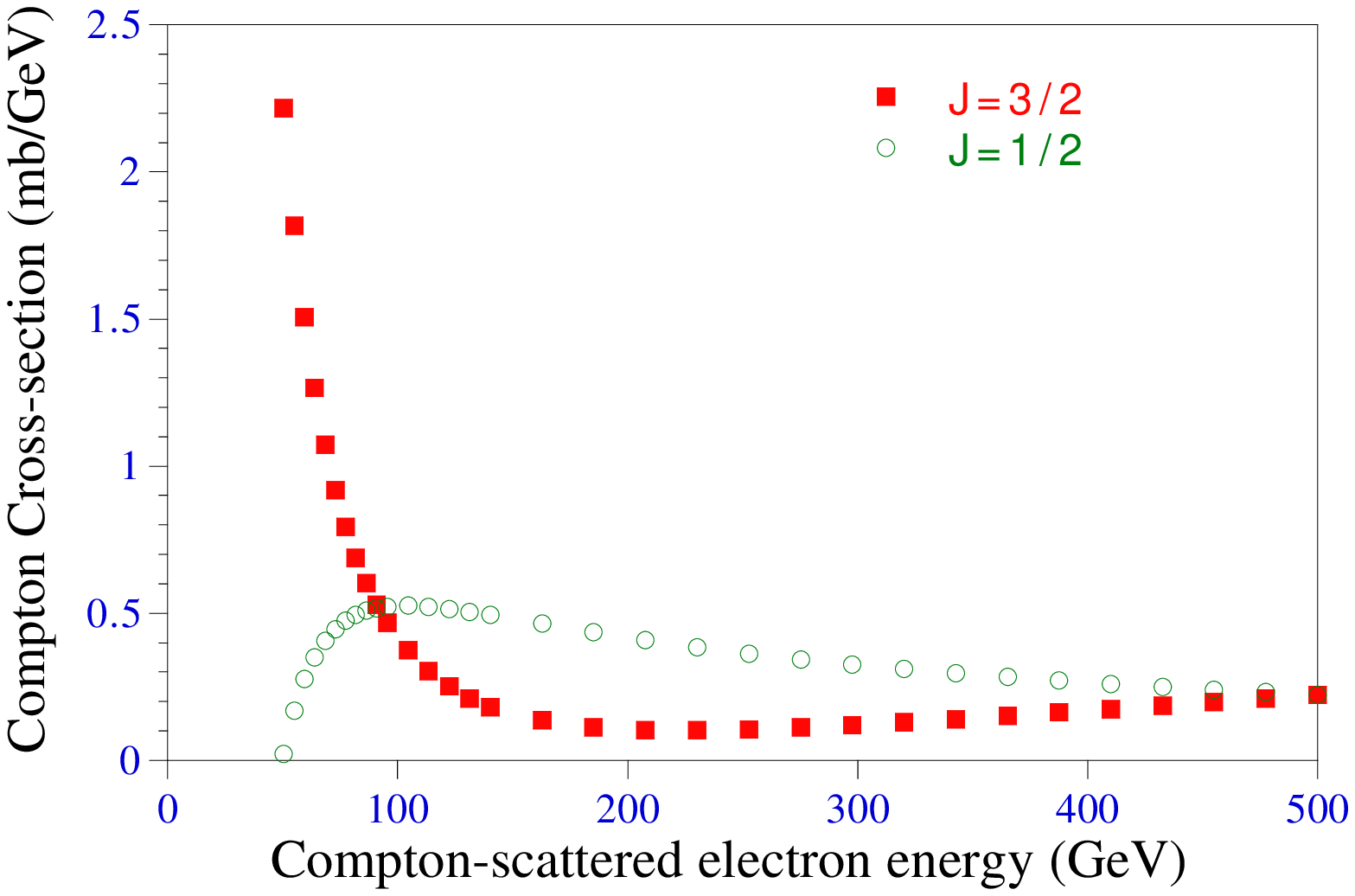,width=6.2cm}
\fcaption{Compton scattering cross-sections for the J=3/2 and J=1/2 
polarization states}
\label{fig:Compxsect}
\end{wrapfigure}

Following the experience of the SLD Compton polarimeter, we expect to measure
the beam polarization primarily from measurements of the Compton rate asymmetry
of the Compton-scattered electrons at the kinematic endpoint. 
The endpoint and endpoint asymmetry are given by the following equations:
\begin{eqnarray}
y & = & (1+\frac{4E_eE_\gamma}{m_e^2})^{-1} \nonumber \\
E_C(endpoint) & = & E_e \cdot y \nonumber \\
A_C(endpoint) & = & \frac{y^2-1}{y^2+1}. \nonumber
\end{eqnarray}   
They are plotted versus the beam energy in Figure~\ref{fig:endpoint} 
for a laser photon energy
of 1.165eV (corresponding to a 1064nm Nd:YAG laser).  At high beam energies,
the Compton endpoint is well separated from the beam energy; this is 
important and helps allow a layout of the extraction line that achieves a
good suppression of background to Compton signal.  The Compton 
asymmetry is also very large, facilitating quick and accurate polarization
measurements.  For the example of a 1.165eV laser photon 
scattering from a 500 GeV beam electron, the Compton cross-sections
for the J=3/2 and J=1/2 polarization states are 
plotted in Figure~\ref{fig:Compxsect}.

\begin{figure}
\begin{center}
\epsfig{figure=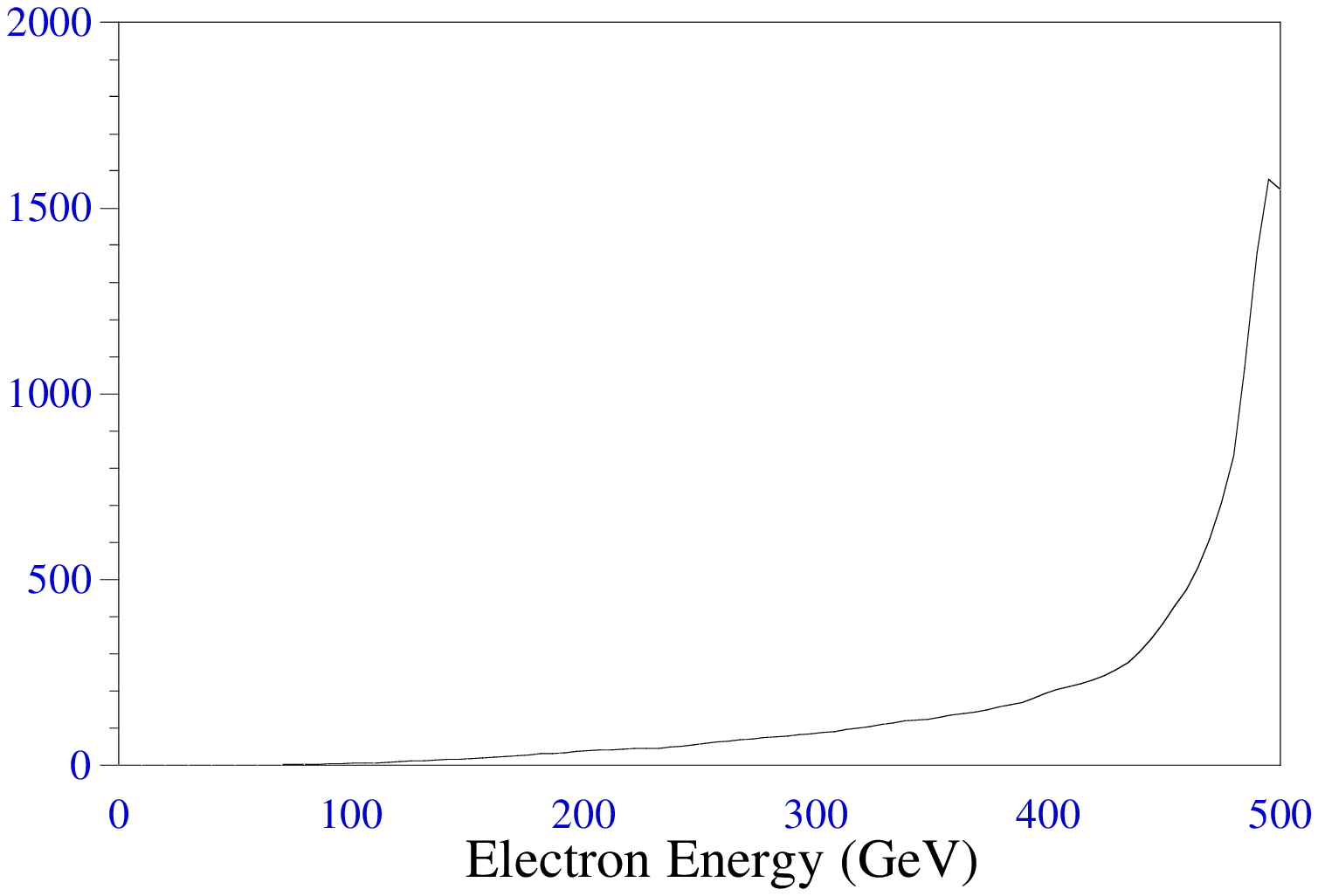,width=6.2cm} \hfill
\epsfig{figure=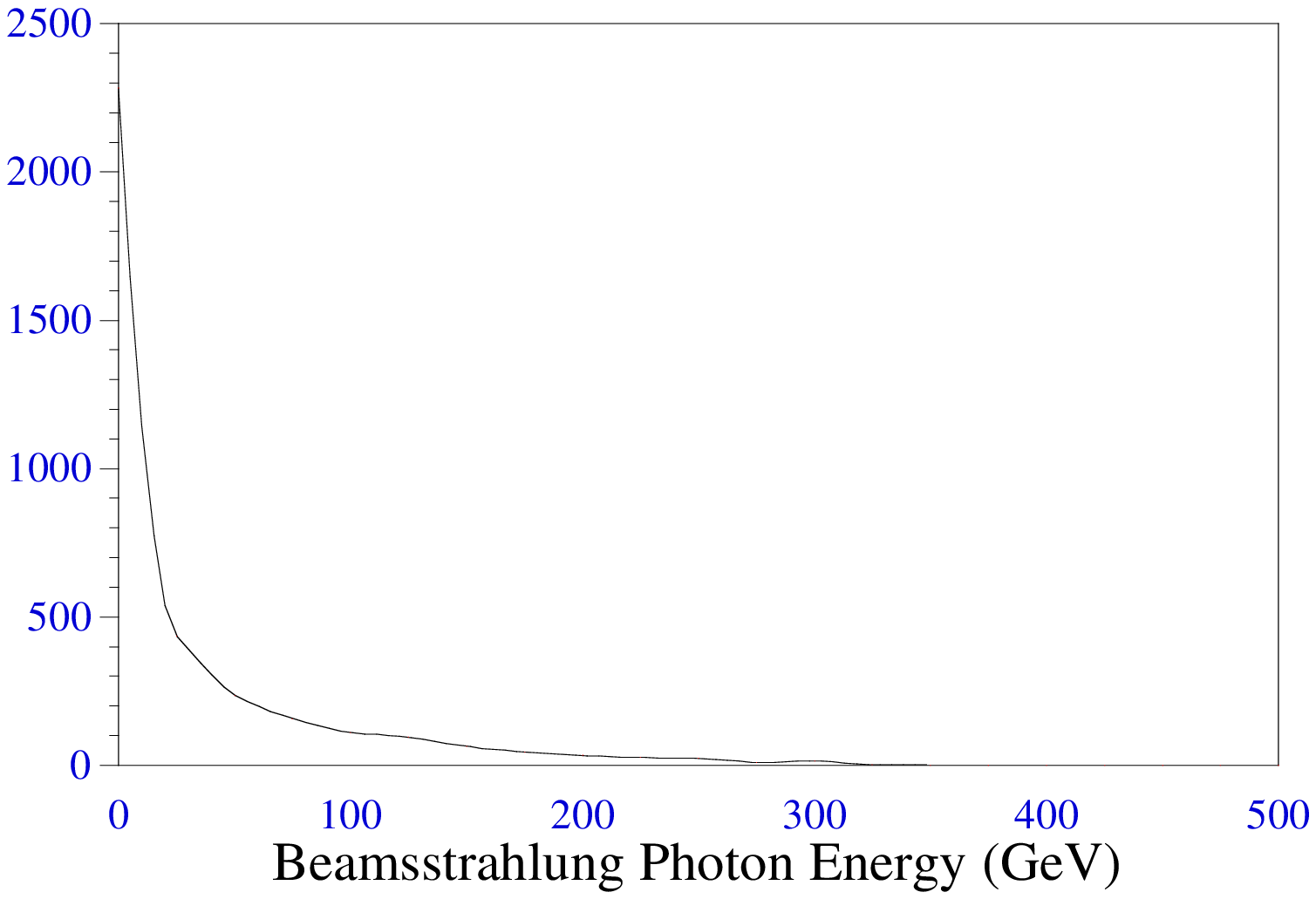,width=6.2cm}
\end{center}
\fcaption{Disrupted electron energy and beamstrahlung energy distributions
at a TLC}
\label{fig:disrupted}
\end{figure}

The primary backgrounds to contend with are collision-related.  The outgoing
beams from the IP are severely disrupted and there is a large flux of 
beamstrahlung.  To illustrate this, we reproduce two plots from SLAC's
Zeroth-order Design Report (ZDR)\cite{ZDR} for the NLC in 
Figure~\ref{fig:disrupted}.  The
average energy loss per incident beam particle is about $10\%$, significantly 
greater than the $0.1\%$ loss at the SLC.  

\begin{figure}
\centering
\epsfig{figure=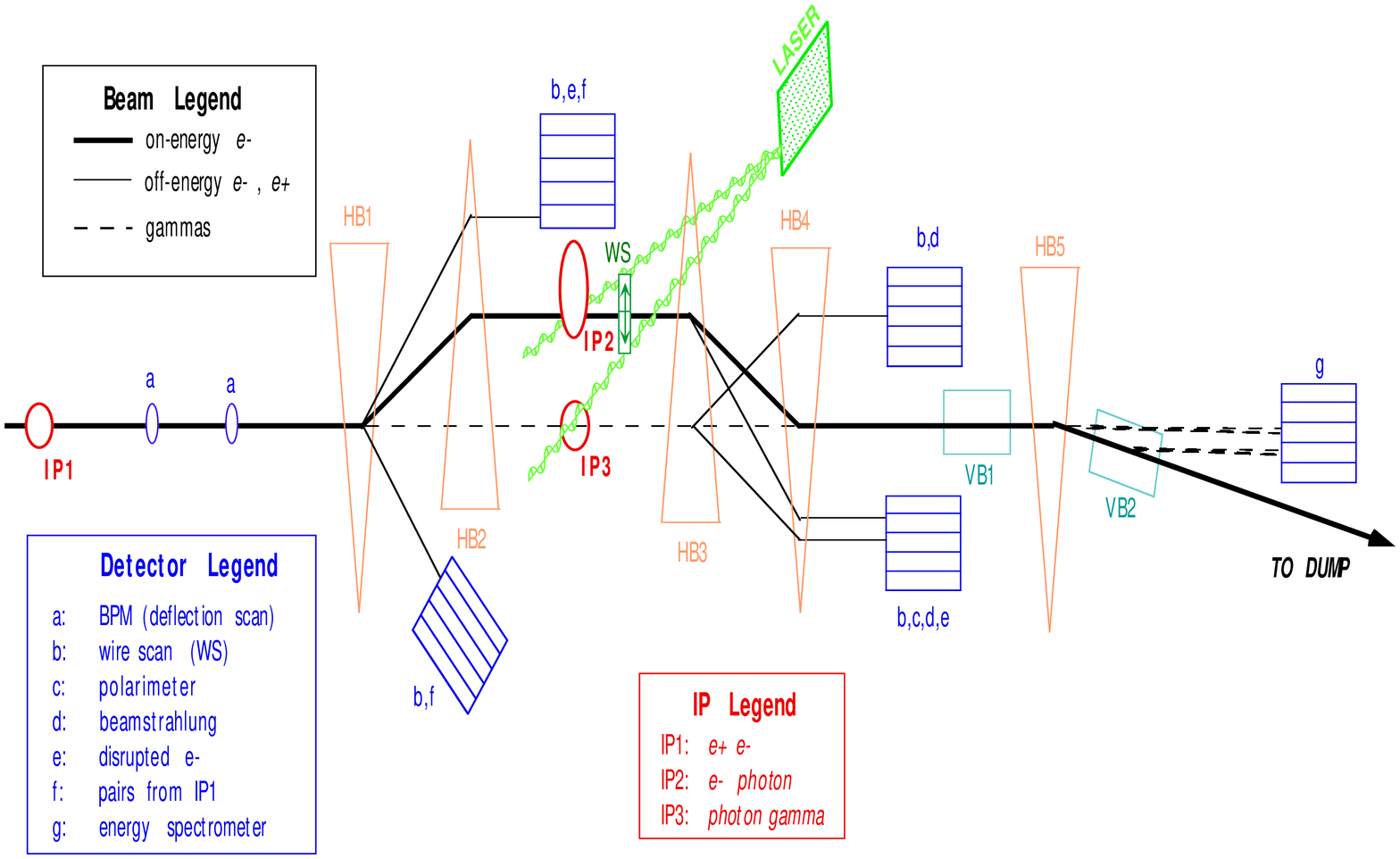,width=12cm}
\fcaption{Extraction Line from IP to Beam Dump}
\label{fig:extline}
\end{figure}

At a TLC, the beam power is roughly 10 MW with a corresponding beamstrahlung
power of 1 MW to be compared with 30 kW beam power and 30 W 
beamstrahlung power at the SLC.  This imposes a challenging environment for 
beam
diagnostics at the TLC.  Very careful design and simulation of the 
extraction lines from the IP to the beam dumps is needed to transport both the
disrupted incident beam and the beamstrahlung with minimal losses, while
allowing for sufficient beam diagnostics.  For SLAC's ZDR a proposed layout
for the extraction line was developed, which is shown in 
Figure~\ref{fig:extline}.

The Compton IP is chosen to be at a location of high dispersion with
$\eta=20$mm.  This assists separating the Compton signal from backgrounds at 
the polarimeter detector.  It also allows studying the dependence of the beam 
polarization on the disrupted electron energy by varying the targeting of
the laser beam on the dispersed electron beam.  This requires a good
measurement of the disrupted energy distribution, which should be achievable 
with a conventional wire scanner.  

The polarization of the incident electron beam prior to colliding is easily 
determined from measurements where the opposing colliding beam is absent.
At the TLC there can be significant depolarization at the level of a few 
percent in the collision process.\cite{Chen}  Though this can be calculated
given a good knowledge of the beam parameters, it is important to measure
the depolarization directly.  By comparing polarization measurements with and 
without collisions and under differing luminosity configurations, it should be 
possible to understand the depolarization loss  to better than $1\%$.

A new feature of the TLC compared to the SLC is the use of bunch trains,
with typically 90 bunches per train and an interbunch spacing of 2.8ns.
It is of interest to measure how the polarization varies within the train.
This can be done by colliding a short ($<2$ns) laser pulse with an individual
electron bunch.  A fast photomultiplier tube and readout gate can then be
used to minimize background from other bunches in the train.  The laser and 
gate timing can be adjusted to map out the polarization within the train.

\section{Conclusions}
A Compton polarimeter located in the extraction line from the IP should achieve
$1\%$ accuracy at a TLC.  It is important to take polarization 
measurements parasitic to beam collisions and detector data logging.  The
outgoing beams from the IP are highly disrupted from the collision process.
This and the high power beamstrahlung produced provide significant 
challenges to designing the extraction line.  This is a more difficult 
problem for the electron-electron collider than for the electron-positron
collider.  Detailed system design and simulations are needed to ensure
adequate signal to background in the polarimeter detector.  Detailed studies
are also needed to evaluate how well the luminosity-weighted
polarization for a given collision process can be determined from the Compton
polarization measurements. 

\nonumsection{References}


\begin{thebibliography}{99}
\bibitem{Comptxsect} S.B. Gunst and L.A. Page, {\it Phys. Rev.} {\bf 92},
	970 (1953).
\bibitem{Woods1} M. Woods, SLAC-PUB-7319, (1996).  Refer also to the theses by 
E. Torrence, SLAC-Report-509 (1997); R. King, SLAC-Report-452 (1994);  
A. Lath, SLAC-Report-454 (1994); and R. Elia, SLAC-Report-429 (1994).
\bibitem{Woods2} M. Woods, SLAC-PUB-7320, (1996); and M. Woods, SLAC-PUB-6694
(1995).
\bibitem{PGC} R.C. Field et al., OREXP 97-04 (1997).
\bibitem{QFC} D. Onoprienko et al., UTKHEP-97-20 (1997).
\bibitem{ZDR} NLC Design Group 
(C. Adolphsen et al.), {\it Zeroth-Order Design
Report for the Next Linear Collider}, SLAC Report 474, (1996).
\bibitem{Chen} K. Yokoya and P. Chen, SLAC-PUB-4692, (1988).
\end{thebibliography}
\end{document}